\documentclass[aps,pra,twocolumn,showpacs,superscriptaddress,floatfix,nofootinbib]{revtex4}
\usepackage{graphicx}
\usepackage{nicefrac}
\usepackage{amsmath}
\usepackage{amsfonts}
\usepackage{amssymb}
\usepackage{amsthm}
\usepackage{epsf}
\usepackage{bm}
\usepackage{bbm}
\usepackage{longtable}

\usepackage{dcolumn}

\newcolumntype{.}{D{x}{}{-1}}

\newcommand{\half}{\frac12}

\newcommand{\balpha}{\bm{\alpha}}
\newcommand{\bnabla}{\bm{\nabla}}

\newcommand{\bfr}{\bm{r}}
\newcommand{\bfk}{\bm{k}}
\newcommand{\bfp}{\bm{p}}

\newcommand{\bfx}{\bm{x}}

\newcommand{\hp}{\hat{\bfp}}

\newcommand{\Za}{{Z\alpha}}

\newcommand{\vare}{\varepsilon}

\newcommand{\SixJ}[6]{
        \left\{
        \begin{array}{ccc}
        #1  & #2  & #3 \\
        #4  & #5  & #6 \\
        \end{array}
        \right\}
        }

\newcommand{\lbr}{\langle}
\newcommand{\rbr}{\rangle}

\begin{document}

\title{Relativistic theory of the double photoionization of helium-like atoms}

\author{Vladimir A. Yerokhin}
\affiliation{Institute of Physics, University of Heidelberg, Philosophenweg
  12, D-69120 Heidelberg, Germany}
\affiliation{GSI Helmholtzzentum f\"ur Schwerionenforschung GmbH, Planckstra{\ss}e 1,
D-64291 Darmstadt, Germany}
\affiliation{Center for Advanced Studies, St.~Petersburg State
Polytechnical University, Polytekhnicheskaya 29,
St.~Petersburg 195251, Russia}

\author{Andrey Surzhykov}
\affiliation{Institute of Physics, University of Heidelberg, Philosophenweg
  12, D-69120 Heidelberg, Germany}
\affiliation{GSI Helmholtzzentum f\"ur Schwerionenforschung GmbH, Planckstra{\ss}e 1,
D-64291 Darmstadt, Germany}

\begin{abstract}
A fully relativistic calculation of the double
photoionization of helium-like atoms is presented. The approach is based on the
partial-wave representation of the Dirac continuum states and accounts for
the retardation in the electron-electron interaction as well as
the higher-order multipoles of the absorbed photon. The electron-electron
interaction is taken into account to the leading order of perturbation theory.
The relativistic effects are shown to become prominent already
for the medium-$Z$ ions, changing the shape and the asymptotic behaviour of
the photon energy dependence of the ratio of the double-to-single
photoionization cross section.
\end{abstract}


\maketitle

\section{Introduction}

The ejection of two electrons caused by absorbtion of a
single photon is one of the fundamental few-body processes in
atomic physics. The process is called double photoionization or, less frequently,
photo-double-ionization. The characteristic feature of this process is that it
proceeds exclusively through the electron-electron interaction. Because
of this, double photoionization has long been used as a testing ground 
for understanding of the electron correlation phenomena.

The traditional system for studying double photoionization is the helium atom, 
for which numerous
experimental and theoretical investigations have been performed during the
last four decades (for a recent review, see  Ref.~\cite{briggs:00}). 
Most widely studied is the ratio of the double-to-single
photoionization cross sections $R = \sigma^{++}/\sigma^+$ as a function of the
energy of the incoming photon $\omega$. Early calculations of this ratio were mainly
concentrated either at the near-threshold region of the photon energies
($\omega \gtrsim \omega_{\rm cr} \approx$ 79 eV for helium, with $\omega_{\rm
  cr}$ being the double ionization energy), where Wannier theory is shown to be
applicable \cite{kossmann:88,maulbetsch:92}, or at the asymptotical nonrelativistic
region $\omega_{\rm cr} \ll \omega \ll m$ \cite{amucia:75,andersson:93,forrey:95}
(where $m$ is the electron mass).
The intermediate region of the photon energies ($\omega
\approx$ 200~eV for helium, where $R(\omega)$ reaches its maximum)
turned out to be much more difficult for an accurate theoretical treatment. Reliable
theoretical predictions in this region
were obtained only in the end of 1990th by means
of sophisticated many-body techniques (notably, the close-coupling methods)
\cite{tang:95,kehifets:96,meyer:97,qiu:98}.

Since the helium atom is an essentially nonrelativistic system, any
relativistic effects in 
its double photoionization are considered to be of little importance at the present
level of experimental precision. However, the recent experiments on the double
$K$-shell photoionization of moderately heavy atoms up to silver
\cite{kantler:99,southworth:03,kantler:06,hoszowska:09,hoszowska:10}
demonstrated significant enhancements (of about factor of five for silver)
of the cross section as compared with results of nonrelativistic
calculations. A natural candidate for explaining this enhancement would be
relativity, which obviously cannot be disregarded when dealing with the deeply
bound electron states in silver.
Because of the relative isolation of the $K$-shell electrons from
the outer electrons, the double $K$-shell ionization of a heavy atom is
often compared to the double photoionization of the corresponding helium-like ion. However, an accurate
theoretical treatment of double $K$-shell ionization should include both the
relativistic effects on the inner-shell electrons and the electron-correlation
effects induced by the outer shells. Such a calculation is rather difficult
and has not been performed so far.

In contrast to many-electron atoms, the helium-like ion is a relatively simple 
system for which an {\it ab initio} description is feasible. So far, there have not been any direct
measurements of the double photoionization of helium-like ions. However, with the
advent of new powerful light sources, such as the free-electron laser (FLASH) in
Hamburg, the Linear Coherent Light Source (LCLS) at Stanford, and the X-ray
Free Electron Laser (XFEL) at Hamburg, the experimental study of various
photoabsorbtion processes is going to be possible for a great variety of ions
in different charge states \cite{epp:07}. Measurements of the double photoionization in
moderately heavy helium-like ions would allow us to effectively
test our understanding of the {\em relativistic} electron-correlation effects.

An alternative approach to the investigation of the double photoionization of
highly charged ions is to study it in the inverse kinematics,
through the radiative double electron capture. For bare oxygen, such 
measurement was recently accomplished in Ref.~\cite{simon:10}. The experimental 
upper bound on this process for bare uranium was reported in Ref.~\cite{bednarz:03}. 

The goal of the present investigation is to perform an {\em ab initio} relativistic
calculation of the double photoionization of a helium-like atom. The electron-electron correlation
(both on the initial and the final states) will be taken into account to the
leading order of the perturbation theory. To the given order of perturbation theory, the
treatment is {\em exact}, i.e., includes all multipoles of the absorbed
photon, the retardation in the electron-electron interaction as well as the
interaction of the electrons with the nucleus without any expansion in the
binding field. The treatment is gauge invariant, both with respect to the
gauge used for the absorbed photon and the gauge of the
electron-electron interaction. The higher-order electron-correlation effects
omitted are estimated by comparing the present numerical
results with the experimental data available for helium.

The remaining paper is organized as follows. In Sec.~\ref{sec:1} we present a short
summary of the relativistic
formulas for the single photoionization, which forms the basis
for our treatment of the double photoionization. In Sec.~\ref{sec:2} we describe the QED
theory of the double photoionization to the lowest relevant order of the perturbation
theory. Sec.~\ref{sec:3} presents details of the calculation.
Numerical results are presented and discussed in Sec.~\ref{sec:4}.

The relativistic units ($\hbar=c=m=1$) are used throughout this paper.

\section{Single photoionization}
\label{sec:1}

In this section we summarize the relativistic formulas for the single
photoionization, as they build the basis for the description of the double
photoionization.

Differential cross section of the photoionization of a one-electron atom
is
\begin{align} \label{1}
d\sigma^{+} = \frac{4\pi^2\alpha}{\omega}\, \bigl| \tau_{fi}^{+}\bigr|^2\,
 \delta(\vare_i+\omega-\vare)\,d \bfp\,,
\end{align}
where $\alpha$ is the fine-structure constant, $\omega$ is the energy of
the absorbed photon, $\vare_i$ is the initial (bound-state) energy, 
$\vare$ and $\bfp$ are the energy and
the momentum of the emitted electron, respectively, $\vare = \sqrt{m^2+\bfp^2}$.
The amplitude of the process $\tau^+_{fi}$ is given by
\begin{align} \label{1a}
\tau_{fi}^+ = \lbr \bfp m| R_{\lambda}| \kappa_i\mu_i\rbr\,,
\end{align}
where $|\kappa_i\mu_i\rbr$ denotes the initial Dirac bound state
with the relativistic angular quantum number $\kappa_i$ and its projection
$\mu_i$, and $|\bfp m\bigr>$ is the wave function of the emitted electron.
The general relativistic expression for the photon absorption operator $R_{\lambda}$ is
\begin{align} \label{1aa}
R_{\lambda} = \balpha\cdot\hat{\bm{u}}_{\lambda}\,
e^{i\bfk\cdot\bfr}\, + G\,\bigl(\balpha\cdot\hat{\bfk}-1 \bigr) \,e^{i\bfk\cdot\bfr}\,,
\end{align}
where $\balpha$ is a three-component vector of the Dirac matrices,
$\hat{\bm{u}}_{\lambda}$ is the polarization vector of the absorbed
photon, $\bfk$ is the photon momentum, $\hat{\bfk} = \bfk/|\bfk|$,
and $G$ is the gauge parameter. In our treatment, all
electron states are the eigenfunctions of the Dirac Hamiltonian, so the
gauge-dependent part of $R_{\lambda}$ vanishes identically. However, we keep
the general gauge of the absorbed photon in actual calculations in order to
check the numerical procedure. 

The wave function
$|\bfp m\bigr> \equiv |\bfp m\bigr>_-$ in Eq.~(\ref{1a})
is the one-electron continuum Dirac state with the
asymptotic momentum $\bfp$, helicity $m = \pm \nicefrac12$, and the ``$-$''
asymptotic behaviour (i.e., the plane wave modified by the Coulomb
logarithmic phase plus the incoming spherical wave)
\cite{eichler:95:book}
\begin{align} \label{2}
|\bfp m\bigr>_- = \frac{u(\bfp,m)}{(2\pi)^{3/2}} &\ \,e^{i[pz-\eta\,\ln
    p(r-z)]}
 \nonumber \\ &
+
  f(\theta,\phi)\,\frac{e^{-i[pr+\eta\,\ln 2pr]}}{r}\,
\end{align}
as $|r-z|\to\infty$, where $\eta = Z\alpha\,\vare/p$ is the Sommerfeld parameter.
The free-electron 4-spinors $u(\bfp, m)$ \cite{rose:61} are normalized by
the condition  $\overline{u}\,u \equiv u^+\gamma_0 u = 1$. The explicit
expression of the electron
wave function is
\begin{align} \label{3}
|\bfp m\bigr>_- = \frac1{\sqrt{p\,\vare}}\sum_{\kappa\mu} i^l\,e^{-i\Delta_l}\,
 C_{lm_l,\half m}^{j\mu}\,Y^*_{lm_l}(\hp)\,|\vare\kappa\mu\rbr\,,
\end{align}
where $\Delta_l$ are the scattering phases \cite{eichler:07:review},
$j = |\kappa|-\nicefrac12$, $l = |\kappa+\nicefrac12|-\nicefrac12$, and
$|\vare\kappa\mu\rbr$ are Dirac continuum states with given relativistic
angular momentum $\kappa$ and the momentum projection $\mu$, normalized on the
energy scale \cite{eichler:95:book}.

Taking into account that $d\bfp = p\,\vare\,d\vare d\Omega$,
the single differential cross section of the photoionization of a
hydrogen-like atom is written as
\begin{align}
\frac{d\sigma^{+}}{d\Omega} = \frac{4\pi^2\alpha}{\omega}\,p\,\vare\, \bigl|
\tau_{fi}^{+}\bigr|^2\,,
\end{align}
where the energy of the emitted electron is fixed by the energy conservation,
$\vare = \vare_i+\omega$.

In our analysis of the double photoionization, we will need the cross section
of the single photoionization of a {\em helium-like} atom. In the
independent-particle approximation, the wave function of the initial
two-electron state is
\begin{align} \label{4}
|J_0M_0\bigr> = &\ N \sum_{\mu_a\mu_b}
 C^{J_0M_0}_{j_a \mu_a\,j_b\mu_b}
 \nonumber \\ & \times
 \frac1{\sqrt{2}}\bigl(|\kappa_a\mu_a\bigr>|\kappa_b\mu_b\bigr>
-|\kappa_b\mu_b\bigr>|\kappa_a\mu_a\bigr>\bigr)\,,
\end{align}
where
$N = 1/\sqrt{2}$ for the equivalent electrons and $N=1$ otherwise.
Averaging the cross section over the momentum projection of the initial
state $M_0$, employing the explicit
expression for the continuum Dirac state (\ref{3}), and
integrating over the angles over the emitted electron, we obtain the total
photoionization cross section of a helium-like atom as
\begin{align}
\sigma^+ = &\ \frac{4\pi^2\alpha}{\omega}\,\biggl[\frac1{2j_a+1}\,
 \sum_{\kappa\mu\mu_a} \left|
 \lbr \vare_1\kappa\mu \left| R_{\lambda}(\omega)\right|\kappa_a\mu_a\rbr
\right|^2
 \nonumber \\ &
+ \frac1{2j_b+1}\,
 \sum_{\kappa\mu\mu_b} \left|
 \lbr \vare_2\kappa\mu \left| R_{\lambda}(\omega)\right|\kappa_b\mu_b\rbr
\right|^2\biggr]\,,
\end{align}
where $\vare_1 = \vare_a+\omega$ and $\vare_2 = \vare_b+\omega$
and we assumed that the energy of the photon is sufficient to ionize any of the two
initial-state electrons. If this is not the case, only one of the two terms
in the brackets should be retained.

The electron correlation effects on $\sigma^+$ omitted in the independent-particle
approximation are well studied in the literature. In particular, the leading
nonrelativistic term of the $1/Z$ expansion of the high-energy asymptotics of 
$\sigma^+$ was obtained in Ref.~\cite{mikhailov:06}. However, since we are
presently interested in the ratio of the double-to-single photoionization
cross sections $\sigma^{++}/\sigma^{+}$, we prefer to treat
both these cross sections on the same footing, i.e., to the 
leading nonvanishing order of perturbation theory with respect to the
electron-electron interaction.

\section{Double photoionization: general formulas}
\label{sec:2}

%
%
\begin{figure}[tb]
  \centerline{\includegraphics[width=\columnwidth]{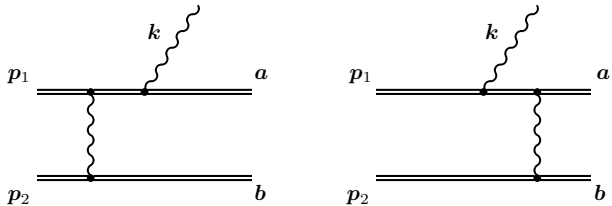}}
\caption{Feynman diagrams representing the double photoionization of a
  helium-like atom. $a$ and $b$ denote the bound electron states, $\bm{p}_1$ and
  $\bm{p}_2$ are the continuum electron states, $\bm{k}$ denotes the incoming
  photon. Double lines denote electrons propagating
  in the binding nuclear field.
  \label{fig:dpi} }
\end{figure}

According to the general rules of quantum field theory \cite{shabaev:02:rep},
the differential cross section of the double photoionization is
\begin{align} \label{e222}
d\sigma^{++} = \frac{4\pi^2\alpha}{\omega}\, \left| \tau_{fi}^{++}\right|^2\,
 \delta(\vare_i+\omega-\vare_1-\vare_2)\,d \bfp_1\, d\bfp_2\,,
\end{align}
where $(\vare_1,\bfp_1)$ and $(\vare_2,\bfp_2)$ are the energy and the
momentum of the two final-state electrons, respectively, $\vare_i$ is the energy of the
initial bound state, and $\tau_{fi}^{++}$ is the
amplitude of the process. To the leading order of perturbation theory, the
amplitude of the process is represented by the two Feynman diagrams shown in
Fig.~\ref{fig:dpi}, where the antisymmetrization of the initial and
the final two-electron states is assumed.

The initial-state wave function is the same as for the single photoionization 
[see Eq.~(\ref{4})] and the initial-state energy is $\vare_i = \vare_a+\vare_b$.
The wave function of the final state is
\begin{align}
|\bfp_1m_1,\bfp_2m_2\bigr>_- = &\
 \frac1{\sqrt{2}}\left(|\bfp_1m_1\bigr>_-|\bfp_2m_2\bigr>_-
 \right. \nonumber \\ & \left.
-|\bfp_2m_2\bigr>_-|\bfp_1m_1\bigr>_- \right)\,.
\end{align}
Note that in the present approach, we use the wave
functions that do not include any electron-correlation effects
(apart from the antisymmetrization); the electron-electron 
interaction enters explicitly into the amplitude of the process
[see Eq.~(\ref{e2}) below]. In the nonrelativistic case, this
approach has been successfully applied to the double
photoionization by many authors, notably in 
Refs.~\cite{amucia:75,mikhailov:04}. Although the formal parameter of the
perturbative expansion is $1/Z$, the leading-term approximation was shown to
adequately describe the double photoionization even for helium. 
Our present treatment is aimed primarily at the helium-like ions, for which
the perturbative expansion converges much faster than for helium. 

The triple differential cross section of the double photoionization obtained
from Eq.~(\ref{e222}) is
\begin{align}  \label{e123}
\frac{d^3\sigma^{++}}{d\Omega_1 d\Omega_2 d\vare_1} = &\
 \frac{4\pi^2\alpha}{\omega}\, p_1p_2\vare_1\vare_2\,
 \nonumber \\ & \times
\sum_{m_1 m_2}
    \left| \tau^{++}_\lambda(\bfp_1m_1,\bfp_2m_2;\omega,J_0M_0)\right|^2\,,
\end{align}
where the energy of the second electron is fixed by the energy conservation,
$\vare_2 = \vare_a+\vare_b+\omega-\vare_1$. The energy distribution of the
emitted electrons can be conveniently parameterized by the
fractional energy sharing parameter $x$,
\begin{align} \label{sharing1}
\vare_1 &\ = m+ x\, (\omega-\omega_{\rm cr})\,, \\
\vare_2 &\ = m+ (1-x) (\omega-\omega_{\rm cr})\,,
\label{sharing2}
\end{align}
where $\omega_{\rm cr}$ is the threshold value of the photon energy 
(the double ionization energy). In our
approximation, $\omega_{\rm cr} = 2m-\vare_a-\vare_b$.

Let us now consider the {\em single differential cross section} $d\sigma^{++}/
d\vare_1$. Substituting the explicit expression for the continuum
wave function (\ref{3}) into Eq.~(\ref{e123}), integrating over the angles, and summing
over $m_1$ and $m_2$, we arrive at
\begin{align}  \label{e124}
\frac{d\sigma^{++}}{d\vare_1} =
 \frac{4\pi^2\alpha}{\omega}\,\sum_{\kappa_1\kappa_2\mu_1\mu_2}
    \left| \tau^{++}_\lambda(\vare_1\kappa_1\mu_1,\vare_2\kappa_2\mu_2;\omega,J_0M_0)\right|^2\,,
\end{align}
where the amplitude contains only the spherical-wave Dirac continuum states.

The {\em total cross section} is obtained as an integral of the single
differential cross section over a half of the energy sharing interval
\begin{align} \label{eq:total}
\sigma^{++} &\ = \int_{m}^{m+(\omega-\omega_{\rm cr})/2}d\vare_1\,
\frac{d\sigma^{++}}{d\vare_1}
  \nonumber \\ &
  = (\omega-\omega_{\rm cr})\, \int_0^{\frac12}dx\,
\frac{d\sigma^{++}}{d\vare_1}\,.
\end{align}
The other half of the energy interval corresponds to interchanging
the first and the second electron, which is already accounted for by the wave
function.

The general expression for the transition amplitude of the double
photoionization process can be obtained by the
two-time Green's function method \cite{shabaev:02:rep},
\begin{widetext}
\begin{align}  \label{e2}
 \tau^{++}_{\lambda}(\vare_1\kappa_1\mu_1,\vare_2\kappa_2\mu_2;\omega,J_0M_0)
  &\ =
 N\,\sum_{\mu_a\mu_b} C^{J_0M_0}_{j_a \mu_a\,j_b\mu_b}\,
\sum_{PQ}(-1)^{P+Q}\, 
 \nonumber \\
&\  \times
\sum_n \biggl\{
\frac{\lbr P\vare_1\,P\vare_2|I(\Delta_{P\vare_2\,Qb})|n\,Qb\rbr\, \lbr n|
  R_{\lambda}|Qa\rbr} {\vare_{Qa}+\omega-\vare_n(1-i0)}
 +
\frac{\lbr P\vare_1| R_{\lambda}|n\rbr\,
 \lbr n\,P\vare_2|I(\Delta_{P\vare_2\,Qb})|Qa\,Qb\rbr}
     {\vare_{P\vare_1}-\omega-\vare_n(1-i0)}
 \biggl\}\,.
\end{align}
\end{widetext}
The first term in the brackets corresponds to the
electron-electron interaction modifying the final-state electron wave
function (the left graph in Fig.~\ref{fig:dpi})
and the second one, the initial-state electron wave function
(the right graph in Fig.~\ref{fig:dpi}).
The summation over $P$ and $Q$  corresponds to the
permutation of the initial and final electrons, $P\vare_1P\vare_2 =
(\vare_1\vare_2)$ or $(\vare_2\vare_1)$, $QaQb = (ab)$ or $(ba)$,
and $(-1)^P$ and $(-1)^Q$ are the permutations sign. The summation over $n$  in Eq.~(\ref{e2})
runs over the complete Dirac spectrum, $\Delta_{ab} \equiv \vare_a-\vare_b$, and
$I(\Delta)$ is the electron-electron interaction operator,
\begin{align}
I(\omega) = e^2 \alpha_{\mu}\alpha_{\nu}\,D^{\mu\nu}(\omega,\bfx_{12})\,,
\end{align}
where $D^{\mu\nu}$ is the photon propagator,
$\alpha_{\mu} = (1,\balpha)$ are the
Dirac matrices. In the Feynman gauge, the
electron-electron interaction takes the form
\begin{align} \label{fey}
I^{\rm Feyn}(\omega) = \alpha\,
(1-\balpha_1\cdot\balpha_2)\,\frac{e^{i|\omega|x_{12}}}{x_{12}}\,,
\end{align}
where $x_{12} =
|\bfx_1-\bfx_2|$.
In the Coulomb gauge, the electron-electron interaction acquires an additional
term, which can be expressed as
\begin{align} \label{cou}
I^{\rm Coul}(\omega)&\  = I^{\rm Feyn}(\omega)
\nonumber \\ &
+ \alpha\,
 \left[ 1-
   \frac{(\balpha_1\cdot\bnabla_1)(\balpha_2\cdot\bnabla_2)}{\omega^2}
\right]\,
\frac{1-e^{i|\omega|x_{12}}}{x_{12}}\,.
\end{align}

\section{Numerical calculation}
\label{sec:3}

The final formulas for the single and double photoionization cross
sections are presented in Appendices~\ref{app:1} and \ref{app:2}.
The calculation of the single photoionization cross section is
straightforward, in contrast to that of the double photoionization.
The major difficulty in the numerical evaluation 
of the double photoionization is the summation over the
complete spectrum of the Dirac equation. In this work we use the approach
based on the analytical representation of the Dirac Coulomb Green function,
which is represented by an infinite sum over the partial waves. The numerical approach was
developed in the previous works \cite{shabaev:00:rec,yerokhin:00:recpra,yerokhin:10:rrec},
where it was used for calculating the QED and electron-electron
interaction corrections to the radiative recombination of electrons with 
highly charged ions.

For a given value of the relativistic angular momentum quantum number $\kappa$,
the radial part of the Dirac Coulomb Green function is represented in terms
of the two-component solutions of the radial Dirac equation
regular at the origin $\left(\phi_{\kappa}^{0}\right)$
and the infinity $\left(\phi_{\kappa}^{\infty}\right)$,
\begin{align}\label{gr01}
 G_{\kappa}(E,r_1,r_2) = &\,
 -\phi_{\kappa}^{\infty}(E,r_1)\,\phi_{\kappa}^{0^T}(E,r_2)\,\theta(r_1-r_2)
\nonumber \\ &
 -\phi_{\kappa}^{0}(E,r_1)\,\phi_{\kappa}^{{\infty}^T}(E,r_2)\,\theta(r_2-r_1)\,,
\end{align}
where $E$ denotes the energy argument of the Green function, $r_1$ and $r_2$
are the radial arguments, and $\theta$ is the step function. For the point
Coulomb potential, the regular and irregular solutions of the radial Dirac
equation are expressed analytically in terms of the Whittaker
functions of the first and second kind
(for explicit formulas see, e.g., Ref.~\cite{mohr:98}).

When the energy argument $E$ is real and greater than the electron rest mass $E>m$,
the Dirac Green function is a complex multi-valued function. So, care should
be taken in this case to choose the correct branch (i.e., the sign of the
imaginary part) of the Green function. The branch of the Green function is
fixed by the sign of the infinitesimal imaginary addition $i0$ in the energy
denominators of Eq.~(\ref{e2}). Specifically, the energy argument
of the Green function in the first term in the brackets of Eq.~(\ref{e2}),
$\vare_{Qa}+\omega$, is greater than the electron mass. For positive-energy
intermediate states, the energy denominator takes the form
$\vare_{Qa}+\omega-\vare_n+i0$, which implies that the energy argument of the
Green function has a small positive imaginary addition and, therefore, the
branch cut of the Green function $[m,\infty)$ should be approached from above.

A serious problem arises in the numerical evaluation of the radial integrals
for the left graph in Fig.~\ref{fig:dpi} (with the electron correlation
modifying the final-state wave function). In this case, the
continuum-state Dirac wave function has to be integrated together with the
Dirac Green function with the energy argument $E>m$ and the spherical Bessel
function. All three functions are strongly oscillating and slowly decreasing
for large radial arguments. It is practically impossible to accurately
evaluate such integral by a straightforward numerical integration. In this
work, we use the method of the complex-plane rotation of the integration
contour, which was previously applied by us to the evaluation of
the free-free transition integrals in the bremsstrahlung
\cite{yerokhin:10:bs}.

We now consider the evaluation of the problematic radial integrals in more
details. Let us introduce the radius of the atom ${\cal R}$, which is defined as
the smallest distance at which all the bound-state electron wave functions
vanish. In the inner region $r<{\cal R}$, evaluation of each partial-wave expansion
term of the amplitude [Eq.~(\ref{e2})]
involves a three-dimensional radial integration. In the
outer region $r> {\cal R}$, however,
all the integrals involving the bound-state wave functions reach
their asymptotical values and only an
one-dimensional integral of the free-free type
needs to be evaluated. The general form of such integral is
\begin{align} \label{e5}
J_{\cal R} = \int_{\cal R}^{\infty}dr\, r^2\, j_L(k_3 r)\, f^i_{\kappa_2}(E_2,r)\,
  \phi_{\kappa_1}^{{\infty}^j}(E_1,r)\,,
\end{align}
where $j_L$ is the spherical Bessel function originating from the photon
propagator in the electron-electron interaction, $f^i$ is the radial
component of the Dirac continuum wave function $(i = 1,2)$, and
$\phi^{{\infty}^j}$ is the radial component of the irregular solution of the
Dirac equation originating from the Green function $(j = 1,2)$. The energy
conservation requires that $E_1 = E_2+k_3$, which leads to the
inequality $p_1 > p_2+k_3$, where $p_{1,2} = \sqrt{E_{1,2}^2-m^2}$ are the
electron momenta associated with the corresponding energies. We now
analytically continue the integrand of Eq.~(\ref{e5}) into the complex $r$
plane and take into account the asymptotical behaviour of individual functions
for large values of $\mbox{\rm Im}(r)$,
\begin{align}
f_{\kappa_2}(E_2,r) &\ \sim \exp\left[p_2\left|\mbox{\rm Im}(r)\right|\right]\,, \\
\phi_{\kappa_1}^{{\infty}}(E_1,r) &\ \sim \exp\left[ p_1\, \mbox{\rm Im}(r)\right]\,, \\
j_l(k_3 r) &\ \sim \exp\left[k_3\,\left|\mbox{\rm Im}(r)\right|\right]\,,
\end{align}
where only the leading exponential behaviour is kept. We observe that, rotating
the integration contour in the integral (\ref{e5})
into the lower complex half-plane $r\to -ir$,
we transform the strongly oscillating integrand into an exponentially decreasing
one, which falls off as $\exp\left[-(p_1-p_2-k_3)\,r \right]$ for large $r$.
After the rotation of the contour, the integral can be easily evaluated
numerically up to a desired precision.

After the radial integrals are successfully evaluated, the remaining problem
is the summation of the partial-wave expansions. Altogether there are five
partial-wave expansions to be delt with: the expansions of the both final
electron wave functions, the Green function, the wave
function of the absorbed photon, and the photon propagator of the electron-electron
interaction. After the angular momentum selection rules are taken into
account, only two partial-wave expansions of the two final-state electrons
remain unbound. The corresponding expansion parameters are
$\kappa_1$ and $\kappa_2$ in Eq.~(\ref{e124}). The convergence of the
resulting double
partial-wave expansion is good for the photon energies near the threshold
$\omega \sim \omega_{\rm cr}$ and
a non-symmetric energy sharing $x\ll\nicefrac12$ but gradually deteriorates
when $\omega$ increases and $x$ approaches $\nicefrac12$. 
In the most difficult case considered here,
$\omega/\omega_{\rm cr} \approx 30$ and $x = \nicefrac12$, up to
$(|\kappa_1|,|\kappa_2|) = (15,15)$ partial waves were included into the calculation.
The truncated partial wave expansion was extrapolated into infinity by using
the $\epsilon$ resummation algorithm (see, e.g., Ref.~\cite{caliceti:07}).

In order to check the numerical procedure, we evaluated the nonrelativistic limit
of our calculations. The easiest way to do this is to decrease
the value of the fine structure constant (or, equivalently, to increase the
value of the speed of light) by a large factor. However, a straightforward
implementation of this scheme leads to numerical instabilities, since
the relativistic operators involve products of the upper and the lower
components of the Dirac wave function and the lower component vanishes in the
nonrelativistic limit. In order to avoid this problem, we express
the lower component in terms of the upper component as
\begin{align}
  f_{\kappa}(r) = \frac1{2m}\left( \frac{d}{dr}+\frac{1+\kappa}{r}\right)\,g_{\kappa}(r)\,,
\end{align}
which is valid to the leading order in $\Za$. After this substitution, our
relativistic code yielded a stable numerical limit when the fine
structure constant was decreased by orders of magnitude, thus giving us the
nonrelativistic limit.

\section{Results and discussion}
\label{sec:4}

We start this section with a discussion of
the total cross section of the double photoionization.
As explained previously, our approach accounts for
the electron correlation to the first order of perturbation theory. Within
this approximation we perform a rigorous relativistic treatment, without
any further simplifications. This approach is expected to yield accurate
results for heavy helium-like ions, whereas for light ions, the precision
of the method
gradually deteriorates with decrease of the nuclear charge. For helium,
the relativistic effects are weak but the electron correlation is strong,
so that in this case our approach predictably yields worse results
than the modern
nonrelativistic methods. However, we will use the helium
case for estimating the higher-order electron correlation effects omitted,
having in mind that in helium-like ions these effects are suppressed
by the inverse power of the nuclear charge.

\begin{figure}
  \centerline{\includegraphics[width=\columnwidth]{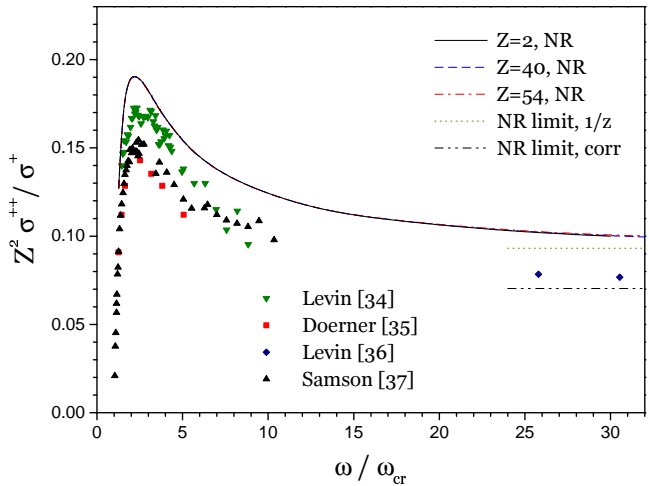}}
\caption{(Color online)
Comparison of the nonrelativistic limit of the
present calculations with the experimental results for
helium and the nonrelativistic theoretical high-energy limits.
The solid (black) line shows our nonrelativistic
results for $Z=2$; the dashed (blue) line, for $Z=40$;
the dashed-dotted (red) line, for $Z=54$. (The three lines are practically
indistinguishable on the picture.) The triangles, squares, and diamonds represent
the experimental results for helium obtained in
Refs.~\cite{levin:96,doerner:96,levin:93,samson:98},
respectively. The dotted line shows the nonrelativistic high-energy
limit calculated to the leading order of the perturbation theory
\cite{amucia:75}. The dash-dot-dotted line shows
the nonrelativistic asymptotical limit calculated with the fully correlated wave
functions \cite{andersson:93,forrey:95}.
  \label{fig:NR} }
\end{figure}

In Fig.~\ref{fig:NR}, we compare the nonrelativistic limit of our calculations
of the ratio of the double-to-single photoionization cross section
$R(\omega) \equiv \sigma^{++}/\sigma^+$ in
helium-like ions with the experimental data available for helium and with the
theoretical high-energy limits. In order to facilitate the analysis of the
results, we plot the scaled ratio $Z^2\,R(\omega)$ as a function of
the ratio $\omega/\omega_{\rm cr}$.

Firstly, we confirm the known statement
\cite{kornberg:94,mikhailov:09} that the {\em nonrelativistic limit} of
the scaled ratio $\overline{R}(\omega)\equiv Z^2 R(\omega/\omega_{\rm cr})$
is well described by a universal function that does not depend on
the nuclear charge number. We will show later that this universal scaling is
strongly violated by the relativistic effects. Fig.~\ref{fig:NR} shows that
the numerical results calculated for different nuclear charges are practically
indistinguishable from each other in the nonrelativistic limit.

Secondly, we
observe good agreement between our nonrelativistic results and 
the leading term of the $1/Z$ expansion of the asymptotics
\cite{amucia:75,mikhailov:98}, $R_{0,1/Z} = 0.0932/Z^2$. The deviation of our
numerical results from the experimental data
\cite{levin:96,doerner:96,levin:93,samson:98} is consistent with the deviation of the
perturbative asymptotical value $R_{0,1/Z}$ from the fully correlated
result $R_{0,\rm corr} = 0.0658/Z^2$ \cite{andersson:93,forrey:95}.
We, therefore, estimate the higher-order electron-correlation effects omitted
in the present treatment to be about $ 2/Z \times 30\%$  for the ratio
$R(\omega)$.

%
%

\begin{figure*}
  \centerline{\includegraphics[width=\textwidth]{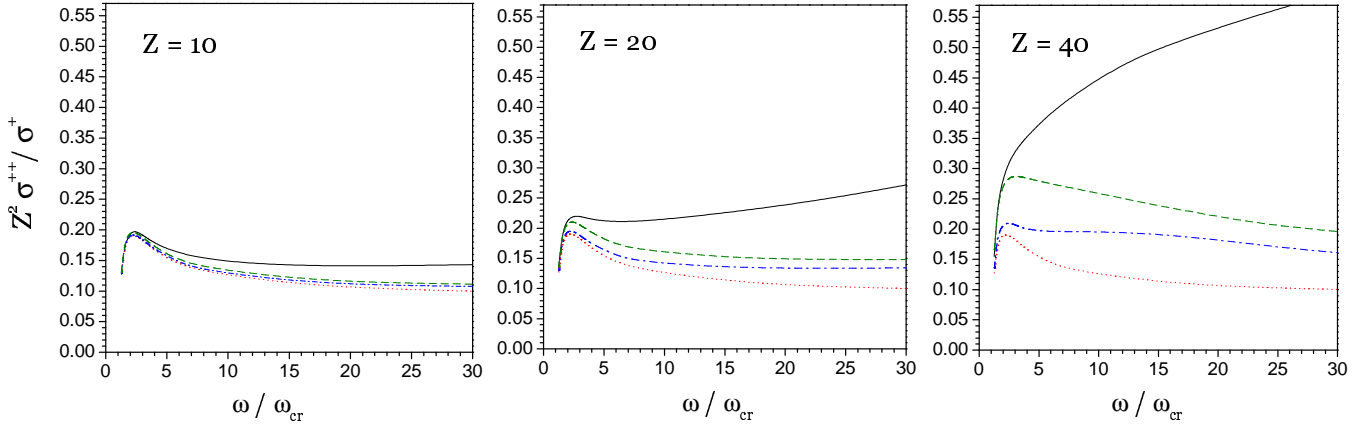}}
\caption{(Color online)
Comparison of the results of the fully relativistic calculations (solid
line, black) with results of various approximate treatments. The dotted (red) line
represents the nonrelativistic limit. The dashed-dotted (blue) line is
obtained with the relativistic wave functions but without retardation and
with including the relativistic $E1$ transition only. The dashed (green) line
shows the results obtained with the relativistic wave functions and the full
retardation but with including the relativistic $E1$ transition only.
The threshold photon energy of the double photoionization is $\omega_{\rm
  cr} = 2.56$~keV for Ne$^{8+}$, $\omega_{\rm  cr} = 10.60$~keV for Ca$^{18+}$, and $\omega_{\rm
  cr} = 43.75$~keV for Zr$^{38+}$.
  \label{fig:diff} }
\end{figure*}

We now turn to our relativistic calculations. Fig.~\ref{fig:diff} presents
our numerical results obtained with a step-by-step inclusion of individual
relativistic effects for several helium-like ions.
The four different treatments compared are (i) the nonrelativistic
calculation (dotted line), (ii) the calculation with the relativistic wave functions but
with neglecting the retardation in the electron-electron interaction and all
multipoles of the absorbed photon higher than the relativistic $E1$
transition (dash-dotted line), (iii) the calculation with the relativistic wave functions and
the full retardation but without the higher multipoles of the absorbed photon
(dashed line),
and (iv) the full relativistic treatment (solid line).

We observe that for moderate photon energies, $\omega/\omega_{\rm cr}
\lesssim 3$, the dominant relativistic effect is the retardation in the
electron-electron interaction, whereas for high photon energies,
$\omega/\omega_{\rm cr} \gtrsim 10$, the effect of the higher multipoles
of the absorbed photon becomes dominant. Altogether, the relativistic effects
are large and change qualitatively the shape of the energy dependence of the
ratio $R(\omega)$ already for medium-$Z$ ions. In particular, the maximum of the curve
located for helium at $\omega/\omega_{\rm cr} \approx 2.5$ disappears for
$Z > 20$ and $R(\omega)$ becomes a monotonically growing function
(at least, up to the maximal photon energies accessible in our calculations),
which is in contrast to the nonrelativistic case where $R(\omega)$ gradually
decreases to approach
a constant high-energy limit from above. It is interesting to note that
similarly strong relativistic effects were recently reported for the nonelastic
electron scattering from the hydrogen-like ions \cite{bostock:09}.
Significant effects caused by the magnetic dipole and electric quadrupole transitions
in the double photoionization of helium-like ions were previously reported
in Ref.~\cite{mikhailov:03}.

In order to check our relativistic calculations, we employed different gauges for the
absorbed photon and, separately, for the photon propagator of the electron-electron interaction.
We found that the gauge-dependent term in the
photon absorbtion operator [see Eq.~(\ref{1aa})] vanishes in the actual calculations.
Independently, we demonstrated that our calculations with the
electron-electron interaction operator in the Feynman and the Coulomb gauges
[see Eqs.~(\ref{fey}) and (\ref{cou})] yield the same results.
This was an important cross-check of the numerical procedure
since the contributions of the two diagrams in
Fig.~\ref{fig:dpi} are separately not gauge invariant.
As an additional test, we calculated the contribution of the right graph in
Fig.~\ref{fig:dpi} independently by a different numerical technique based on
the $B$-spline representation of the Dirac Coulomb Green function 
\cite{shabaev:04:DKB}.

\begin{figure}
  \centerline{\includegraphics[width=\columnwidth]{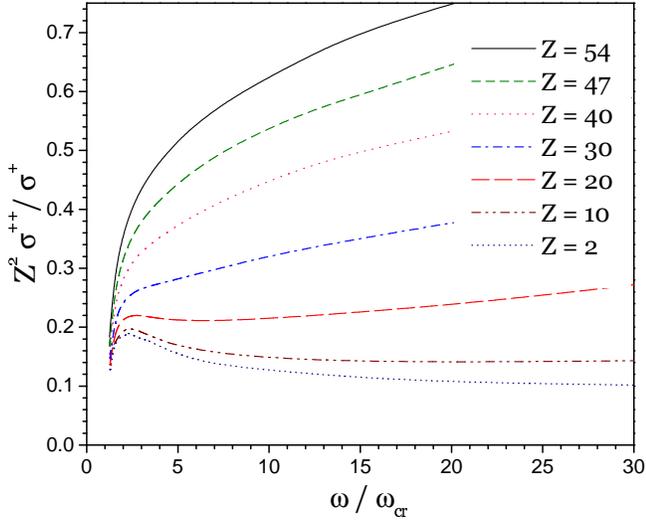}}
\caption{(Color online) Relativistic results for the
scaled ratio of the double photoionization cross
section $\sigma^{++}$ to the single photoionization cross section $\sigma^{+}$,
for helium-like ions with different nuclear charges $Z$, as a function of
the photon energy $\omega$ divided by the threshold photon energy of the
double photoionization $\omega_{\rm cr}$.
  \label{fig:full} }
\end{figure}

\begin{figure}
  \centerline{\includegraphics[width=\columnwidth]{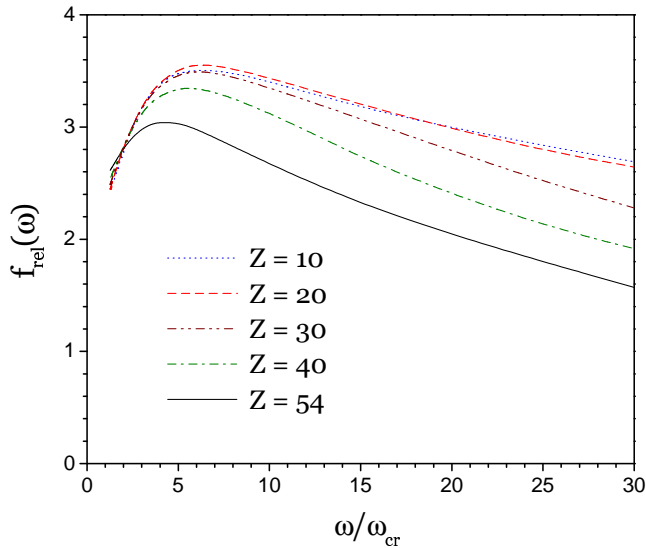}}
\caption{(Color online) Relativistic enhancement function $f_{\rm rel}$
  defined by Eq.~(\ref{relenh}), for
  different helium-like ions.
  \label{fig:relenh} }
\end{figure}

The results of our relativistic calculations for
various helium-like ions are summarized in Fig.~\ref{fig:full}. It can be
readily seen that the relativistic effects strongly violate the nonrelativistic
scaling rule stating that the function $\overline{R}(\omega)\equiv Z^2 R(\omega/\omega_{\rm
  cr})$ does not depend on the nuclear charge. The relativistic enhancement of
the ratio of the double-to-single photoionization cross sections can be conveniently
parameterized as
\begin{align} \label{relenh}
R(\omega,Z) = R_{\rm NR}(\omega,Z)\,\left[ 1 + (\Za)^2\,\frac{\omega}{\omega_{\rm
      cr}}\, f_{\rm rel}(\omega,Z)\right]\,,
\end{align}
where $R_{\rm NR}$ is the nonrelativistic limit of the ratio
$\sigma^{++}/\sigma^+$ and
$f_{\rm rel}$ is a smooth function of the nuclear charge $Z$ and photon
energy $\omega$. This function is plotted for several nuclear charges in
Fig.~\ref{fig:relenh}. We observe that the numerical values of $f_{\rm rel}$ 
lay within the interval of $(1.5,3.5)$ for the wide range of nuclear charge
numbers and photon energies. We also found that for $Z \lesssim 20$,
$f_{\rm  rel}\approx 3$ and is nearly $Z$-independent. In particular, for
helium and 1~keV photon energy ($\omega/\omega_{\rm cr} \approx 13$), the
above formula predicts the relativistic enhancement of about 0.8\%, which is
negligible at the current level of experimental precision.

\begin{figure*}
  \centerline{\includegraphics[width=\textwidth]{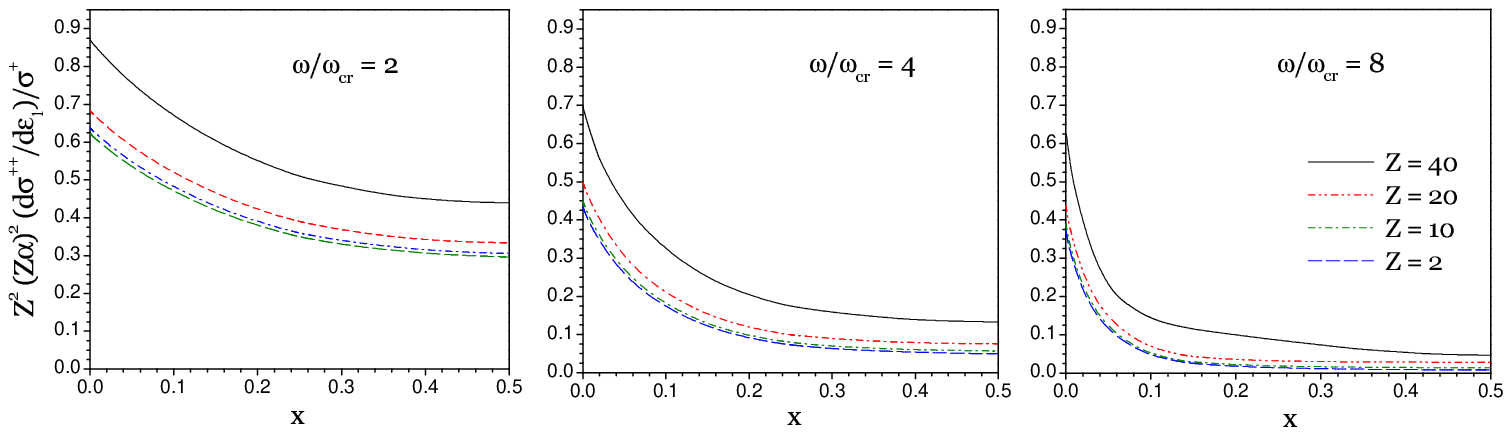}}
\caption{(Color online) Normalized energy distribution of the ejected electrons as a
  function of the energy sharing parameter $x$ defined by
  Eqs.~(\ref{sharing1}) and (\ref{sharing2}),
  for different nuclear charge numbers and photon energies.
  \label{fig:sharing} }
\end{figure*}

The normalized energy distribution of the ejected electrons is shown in
Fig.~\ref{fig:sharing} for several nuclear charges and photon energies. Again,
we observe that the scaled energy distribution does not depend on the nuclear
charge in the nonrelativistic limit and that this scaling is violated by the
relativistic effects. Note that in the case of the energy distribution,
the differential cross section has to be scaled by the factor of
$Z^2(\Za)^2/\sigma^{+}$ and not by $Z^2/\sigma^{+}$ as for the
total cross section; the additional factor is
due to the fact that the integration interval in Eq.~(\ref{eq:total}) is
proportional to $(\Za)^2$. We also confirm the known statement
\cite{amucia:75} that the highly asymmetric energy
sharing between the two emitted electrons
dominates at the high photon energies,
with one of the electrons carrying away most of the photon
energy.

%
%

%
%

%
%

%
%

\section{Conclusion}

In this paper, we presented a relativistic calculation of the double
photoionization of helium-like atoms. Our approach is based on the
partial-wave representation of the Dirac continuum states and accounts for
the retardation in the electron-electron interaction,
the higher-order multipoles of the absorbed photon as well as the
interaction of the electrons with the nucleus without any expansion in the
binding field. The electron-electron interaction is taken into account to the
first order of perturbation theory. The omitted higher-order electron correlation
effects are estimated by comparing our numerical results for atomic
helium with the experimental results and the available nonrelativistic theory.
The calculational results are shown to be gauge invariant both with respect of
the gauge of the absorbed photon and the gauge of the
electron-electron interaction.

Our calculation shows that the relativistic effects become
prominent in the double photoionization cross section
already for medium-$Z$ ions. These
effects change the shape of the energy dependence of the
ratio of the double-to-single photoionization $R(\omega)$ drastically. In particular,
the well-known constant high-energy asymptotic behaviour of $R(\omega)$ in
helium gives place to the monotonically growing behaviour in the case of
helium-like targets with $Z\gtrsim 20$.

We note that the reported magnitude of the relativistic effects is insufficient to
explain the large discrepancy between theory and experiment observed for the
double $K$-shell photoionization in neutral atoms. In particular, for the
photon energy of 90~keV in silver $(\omega/\omega_{\rm cr} = 1.73)$, we
obtain the relativistic enhancement of about 50\% [see
  Eq.~(\ref{relenh})], which is much smaller than the 4-fold disagreement
between theory and experiment \cite{kantler:06}.

It should be mentioned that in the present investigation we do not consider
the Compton scattering mechanism of the total double photoionization, 
which is known to contribute significantly for
high-energy photons \cite{andersson:93}. This reaction channel is more
difficult for a relativistic calculation than the photoabsorption considered
in the present work. To a certain extent, the photoabsorption and the Compton
mechanisms of the double photoionization can be distinguished in the experiment
by measuring the momentum vector of the recoil ions \cite{spielberger:95}.

\section*{Acknowledgement}

The work reported in this paper was supported by the Helmholtz Gemeinschaft
(Nachwuchsgruppe VH-NG-421).
The computations were partly performed on the
computer cluster of St.~Petersburg State Polytechnical University.

\appendix
\section{Single photoionization: calculation formulas}
\label{app:1}

In order to perform integrations over the angular variables in the
matrix element of the
photon absorption operator, we  fix the $z$-axis of our
coordinate system to be directed along the photon momentum $\bfk$ and assume
that the polarization vector $\hat{\bm{u}}$ has the only
nonvanishing spherical component $u_{\lambda}$ ($\lambda = \pm 1$).
The result for the matrix element of the gauge-independent part of the photon absorption operator
is
\begin{widetext}
\begin{align}
\lbr \kappa_a \mu_a|\balpha\cdot\hat{\bm{u}}_{\lambda}\, e^{i\bfk\cdot\bfr} | \kappa_n\mu_n\rbr =
(-1)\sum_{JL}i^{1+L}\,\Pi_L\, (-1)^{j_n-\mu_n}\,
 C_{j_a\mu_a,j_n\,-\mu_n}^{J\lambda}\, C_{L0,1\lambda}^{J\lambda}\, P_{JL}(an)\,,
\end{align}
where $\Pi_L = \sqrt{2L+1}$ and the initial and final states are the Dirac
states with given relativistic angular momentum quantum number $\kappa$ and
the momentum projection
$\mu$. The radial integrals $P_{JL}$ are
\begin{align}
P_{JL}(an) = \int_0^{\infty} dr\, r^2\, j_L(\omega r)\, \bigl[
  g_n(r)\,f_a(r)\,S_{JL}(\kappa_n,-\kappa_a)
 - f_n(r)\,g_a(r)\,S_{JL}(-\kappa_n,\kappa_a)\bigr]\,,
\end{align}
where $j_L$ is the spherical Bessel function,
$g_i$ and $f_i$ are the upper and the lower radial components of the Dirac
wave function, respectively, and
$S_{JL}$ are the angular coefficients
given by Eqs.~(C7)-(C9) of Ref.~\cite{yerokhin:99:pra}.
The matrix element of the gauge-dependent part of the
photon absorption operator can be evaluated as
\begin{align}
\lbr \kappa_a\mu_a |\bigl(\balpha\cdot\hat{\bfk}-1\bigr)\, e^{i\bfk\cdot\bfr} | \kappa_n\mu_n\rbr =
(-1)\sum_{JL}i^{L}\,\Pi_L\, (-1)^{j_n-\mu_n}\,
 C_{j_a\mu_a,j_n\,-\mu_n}^{J0}\, \biggl[ i\,C_{L0,10}^{J0}\, P_{JL}(an)
 + \delta_{J,L}\,C_{L}(\kappa_a,\kappa_n)\,R_L(an)\biggr]\,,
\end{align}
where the angular coefficients $C_L$ are given by Eq.~(C10)
of Ref.~\cite{yerokhin:99:pra} and $R_L$ is the radial integral
\begin{align}
R_{L}(an) = \int_0^{\infty} dr\, r^2\, j_L(\omega r)\,\bigl[
  g_n(r)\,g_a(r) + f_n(r)\,f_a(r)\bigr]\,.
\end{align}
\end{widetext}

\section{Double photoionization: calculation formulas}
\label{app:2}

Let us write the amplitude (\ref{e2}) as
\begin{align}  \label{e2a}
\tau^{++}_{\lambda}  = \tau_{12ab}- \tau_{21ab}-\tau_{12ba}+\tau_{21ba}\,,
\end{align}
where $\tau_{12ab}$ corresponds to the part of Eq.~(\ref{e2}) with
$P\vare_1P\vare_2 =\vare_1\vare_2$ and $QaQb  = ab$ and the remaining three
terms are obtained by permutations. The
contributions due to the two terms in the braces  of Eq.~(\ref{e2})
will be denoted by subscripts
$A$ and $B$, respectively, $\tau_{12ab} = \tau_{12ab,A}+\tau_{12ab,B}$.

The calculation formulas take the simplest form in the case when the initial
electron state is the ground state of the atom. In this case, the permutation
over the $a$ and $b$ electrons yields just a combinatorial factor and
all summations over the momentum projections can be evaluated in the closed
form. The result for the single differential cross section is
\begin{align}
\frac{d\sigma^{++}}{d\vare_1} =
  \frac{4\pi^2\alpha}{\omega}\, \sum_{\kappa_1\kappa_2 J}\,
 \biggl| 2\left[ \widetilde{\tau}_{12ab} + (-1)^{j_1-j_2+J}\,\widetilde{\tau}_{21ab}\right]\biggr|^2\,.
\end{align}
The amplitude is $\widetilde{\tau}_{12ab} =
\widetilde{\tau}_{12ab,A}+\widetilde{\tau}_{12ab,B}$,
\begin{widetext}
\begin{align}
\widetilde{\tau}_{12ab,A} = \frac{\alpha}{\sqrt{2}} \sum_{LL'\kappa_n}
 \, (-1)\,i^{1+L}\,\Pi_L\, C_{L0,\, 1\lambda}^{J\lambda}\,
 \frac{(-1)^{L'+J+j_n+j_2}}{\sqrt{2}}\,\SixJ{j_1}{j_2}{J}{\frac12}{j_n}{L'}
  \sum_n \frac{R_{L'}(\Delta,\vare_1\vare_2nb)\, P_{JL}(na)}{\vare_a+\omega-\vare_n}
\,,
\end{align}
and
\begin{align}
\widetilde{\tau}_{12ab,B} = \frac{\alpha}{\sqrt{2}} \sum_{LL'\kappa_n}
 \, (-1)\,i^{1+L}\,\Pi_L\, C_{L0,\, 1\lambda}^{J\lambda}\,
 \frac{\delta_{j_n,j_2}}{\sqrt{2}(2j_2+1)}
  \sum_n \frac{P_{JL}(\vare_1 n)\,R_{L'}(\Delta,n\vare_2ab)}{\vare_1-\omega-\vare_n}
\,,
\end{align}
\end{widetext}
where $R_L(\Delta,abcd)$ is the relativistic generalization of the Slater
integral for the electron-electron interaction given by Eqs.~(C1)-(C10) of
Ref.~\cite{yerokhin:99:pra} and $\Delta = \vare_2-\vare_b$.


\begin{thebibliography}{10}

\bibitem{briggs:00}
J.~S. Briggs and V.~Schmidt,
\newblock J. Phys. B {\bf 33}, R1 (2000).

\bibitem{kossmann:88}
H.~Kossmann, V.~Schmidt, and T.~Andersen,
\newblock Phys. Rev. Lett. {\bf 60}, 1266 (1988).

\bibitem{maulbetsch:92}
F.~Maulbetsch and J.~S. Briggs,
\newblock Phys. Rev. Lett. {\bf 68}, 2004 (1992).

\bibitem{amucia:75}
M.~Y. Amucia, E.~G. Drukarev, V.~G. Gorshkov, and M.~P. Kazachkov,
\newblock J. Phys. B {\bf 8}, 1248  (1975).

\bibitem{andersson:93}
L.~R. Andersson and J.~Burgd\"orfer,
\newblock Phys. Rev. Lett. {\bf 71}, 50 (1993).

\bibitem{forrey:95}
R.~C. Forrey, H.~R. Sadeghpour, J.~D. Baker, J.~D. Morgan, and A.~Dalgarno,
\newblock Phys. Rev. A {\bf 51}, 2112 (1995).

\bibitem{tang:95}
J.-Z. Tang and I.~Shimamura,
\newblock Phys. Rev. A {\bf 52}, R3413 (1995).

\bibitem{kehifets:96}
A.~S. Kheifets and I.~Bray,
\newblock Phys. Rev. A {\bf 54}, R995 (1996).

\bibitem{meyer:97}
K.~W. Meyer, C.~H. Greene, and B.~D. Esry,
\newblock Phys. Rev. Lett. {\bf 78}, 4902 (1997).

\bibitem{qiu:98}
Y.~Qiu, J.-Z. Tang, J.~Burgd\"orfer, and J.~Wang,
\newblock Phys. Rev. A {\bf 57}, R1489 (1998).

\bibitem{kantler:99}
E.~P. Kanter, R.~W. Dunford, B.~Kr\"assig, and S.~H. Southworth,
\newblock Phys. Rev. Lett. {\bf 83}, 508 (1999).

\bibitem{southworth:03}
S.~H. Southworth, E.~P. Kanter, B.~Kr\"assig, L.~Young, G.~B. Armen, J.~C.
  Levin, D.~L. Ederer, and M.~H. Chen,
\newblock Phys. Rev. A {\bf 67}, 062712 (2003).

\bibitem{kantler:06}
E.~P. Kanter, I.~Ahmad, R.~W. Dunford, D.~S. Gemmell, B.~Kr\"assig, S.~H.
  Southworth, and L.~Young,
\newblock Phys. Rev. A {\bf 73}, 022708 (2006).

\bibitem{hoszowska:09}
J.~Hoszowska, A.~K. Kheifets, J.-C. Dousse, M.~Berset, I.~Bray, W.~Cao,
  K.~Fennane, Y.~Kayser, M.~Kav\ifmmode \check{c}\else
  \v{c}\fi{}i\ifmmode~\check{c}\else \v{c}\fi{}, J.~Szlachetko, and
  M.~Szlachetko,
\newblock Phys. Rev. Lett. {\bf 102}, 073006 (2009).

\bibitem{hoszowska:10}
J.~Hoszowska, J.-C. Dousse, W.~Cao, K.~Fennane, Y.~Kayser, M.~Szlachetko,
  J.~Szlachetko, and M.~Kav\ifmmode \check{c}\else
  \v{c}\fi{}i\ifmmode~\check{c}\else \v{c}\fi{},
\newblock Phys. Rev. A {\bf 82}, 063408 (2010).

\bibitem{epp:07}
S.~W. Epp, J.~R.~C. L\'{o}pez-Urrutia, G.~Brenner, V.~M\"{a}ckel, P.~H. Mokler,
  R.~Treusch, M.~Kuhlmann, M.~V. Yurkov, J.~Feldhaus, J.~R. Schneider,
  M.~Wellh\"{o}fer, M.~Martins, W.~Wurth, and J.~Ullrich,
\newblock Phys. Rev. Lett. {\bf 98}, 183001 (2007).

\bibitem{simon:10}
A.~Simon, A.~Warczak, T.~ElKafrawy, and J.~A. Tanis,
\newblock Phys. Rev. Lett. {\bf 104}, 123001 (2010).

\bibitem{bednarz:03}
G.~Bednarz, D.~Sierpowski, T.~St\"ohlker, A.~Warczak, H.~Beyer, F.~Bosch,
  A.~Br\"auning-Demian, H.~Br\"auning, X.~Cai, A.~Gumberidze, S.~Hagmann,
  C.~Kozhuharov, D.~Liesen, X.~Ma, P.~Mokler, A.~Muthig, Z.~Stachura, and
  S.~Toleikis,
\newblock Nucl. Instrum. Methods {\bf B 205}, 573  (2003),
\newblock 11th International Conference on the Physics of Highly Charged Ions.

\bibitem{eichler:95:book}
J.~Eichler and W.~Meyerhof,
\newblock {\em Relativistic Atomic Collisions},
\newblock Academic Press, San Diego, 1995.

\bibitem{rose:61}
M.~E. Rose,
\newblock {\em {Relativistic Electron Theory}},
\newblock John Wiley \& Sons, NY, 1961.

\bibitem{eichler:07:review}
J.~Eichler and T.~St\"ohlker,
\newblock Phys. Rev. {\bf 439}, 1  (2007).

\bibitem{mikhailov:06}
A.~Mikhailov, A.~Nefiodov, and G.~Plunien,
\newblock Phys. Lett. A {\bf 358}, 211  (2006).

\bibitem{shabaev:02:rep}
V.~M. Shabaev,
\newblock Phys. Rep. {\bf 356}, 119  (2002).

\bibitem{mikhailov:04}
A.~I. Mikhailov, I.~A. Mikhailov, A.~N. Moskalev, A.~V. Nefiodov, G.~Plunien,
  and G.~Soff,
\newblock Phys. Rev. A {\bf 69}, 032703 (2004).

\bibitem{shabaev:00:rec}
V.~M. Shabaev, V.~A. Yerokhin, T.~Beier, and J.~Eichler,
\newblock Phys. Rev. A {\bf 61}, 052112 (2000).

\bibitem{yerokhin:00:recpra}
V.~A. Yerokhin, V.~M. Shabaev, T.~Beier, and J.~Eichler,
\newblock Phys. Rev. A {\bf 62}, 042712 (2000).

\bibitem{yerokhin:10:rrec}
V.~A. Yerokhin and A.~Surzhykov,
\newblock Phys. Rev. A {\bf 81}, 062703 (2010).

\bibitem{mohr:98}
P.~J. Mohr, G.~Plunien, and G.~Soff,
\newblock Phys. Rep. {\bf 293}, 227  (1998).

\bibitem{yerokhin:10:bs}
V.~A. Yerokhin and A.~Surzhykov,
\newblock Phys. Rev. A {\bf 82}, 062702 (2010).

\bibitem{caliceti:07}
E.~Caliceti, M.~Meyer-Hermann, P.~Ribeca, A.~Surzhykov, and U.~Jentschura,
\newblock Phys. Rep. {\bf 446}, 1  (2007).

\bibitem{kornberg:94}
M.~A. Kornberg and J.~E. Miraglia,
\newblock Phys. Rev. A {\bf 49}, 5120 (1994).

\bibitem{mikhailov:09}
A.~I. Mikhailov, A.~V. Nefiodov, and G.~Plunien,
\newblock J. Phys. B {\bf 42}, 231003 (2009).

\bibitem{mikhailov:98}
A.~I. Mikhailov and I.~A. Mikhailov,
\newblock Zh. Eksp. Teor. Fiz. {\bf 114}, 1537 (1998),
\newblock [Sov.~Phys.~JETP {\bf 87}, 833 (1998)].

\bibitem{levin:96}
J.~C. Levin, G.~B. Armen, and I.~A. Sellin,
\newblock Phys. Rev. Lett. {\bf 76}, 1220 (1996).

\bibitem{doerner:96}
R.~D\"orner, T.~Vogt, V.~Mergel, H.~Khemliche, S.~Kravis, C.~L. Cocke,
  J.~Ullrich, M.~Unverzagt, L.~Spielberger, M.~Damrau, O.~Jagutzki, I.~Ali,
  B.~Weaver, K.~Ullmann, C.~C. Hsu, M.~Jung, E.~P. Kanter, B.~Sonntag, M.~H.
  Prior, E.~Rotenberg, J.~Denlinger, T.~Warwick, S.~T. Manson, and
  H.~Schmidt-B\"ocking,
\newblock Phys. Rev. Lett. {\bf 76}, 2654 (1996).

\bibitem{levin:93}
J.~C. Levin, I.~A. Sellin, B.~M. Johnson, D.~W. Lindle, R.~D. Miller,
  N.~Berrah, Y.~Azuma, H.~G. Berry, and D.-H. Lee,
\newblock Phys. Rev. A {\bf 47}, R16 (1993).

\bibitem{samson:98}
J.~A.~R. Samson, W.~C. Stolte, Z.-X. He, J.~N. Cutler, Y.~Lu, and R.~J.
  Bartlett,
\newblock Phys. Rev. A {\bf 57}, 1906 (1998).

\bibitem{bostock:09}
C.~J. Bostock, D.~V. Fursa, and I.~Bray,
\newblock Phys. Rev. A {\bf 80}, 052708 (2009).

\bibitem{mikhailov:03}
A.~I. Mikhailov, I.~A. Mikhailov, A.~N. Moskalev, A.~V. Nefiodov, G.~Plunien,
  and G.~Soff,
\newblock Phys. Lett. A {\bf 316}, 395  (2003).

\bibitem{shabaev:04:DKB}
V.~M. Shabaev, I.~I. Tupitsyn, V.~A. Yerokhin, G.~Plunien, and G.~Soff,
\newblock Phys. Rev. Lett. {\bf 93}, 130405 (2004).

\bibitem{spielberger:95}
L.~Spielberger, O.~Jagutzki, R.~D\"orner, J.~Ullrich, U.~Meyer, V.~Mergel,
  M.~Unverzagt, M.~Damrau, T.~Vogt, I.~Ali, K.~Khayyat, D.~Bahr, H.~G. Schmidt,
  R.~Frahm, and H.~Schmidt-B\"ocking,
\newblock Phys. Rev. Lett. {\bf 74}, 4615 (1995).

\bibitem{yerokhin:99:pra}
V.~A. Yerokhin and V.~M. Shabaev,
\newblock Phys. Rev. A {\bf 60}, 800  (1999).

\end{thebibliography}

\end{document}